\documentclass[runningheads]{llncs}
\usepackage{hyperref}
\usepackage[dvipsnames]{xcolor}
\usepackage[colorinlistoftodos,prependcaption,textsize=tiny]{todonotes}
\usepackage{graphicx}
\usepackage{microtype}
\usepackage[utf8]{inputenc}
\usepackage{algorithm}
\usepackage[noend]{algpseudocode}
\usepackage{tabularx}
\usepackage{booktabs}
\usepackage{paralist}
\usepackage{etoolbox}
\usepackage{caption}
\usepackage{subcaption}

\hypersetup{
	pdftoolbar=true,        
	pdfmenubar=true,        
	pdffitwindow=false,     
	pdfstartview={FitH},    
	pdftitle={Complex Event Processing under Constrained Resources},    
	pdfauthor={},     
	pdfsubject={},   
	pdfcreator={},   
	pdfproducer={}, 
	pdfkeywords={}, 
	pdfnewwindow=true,      
	colorlinks=true,       
	linkcolor=Brown,          
	citecolor=OliveGreen,        
	filecolor=magenta,      
	urlcolor=NavyBlue           
}

\makeatletter
\providecommand{\subtitle}[1]{
	\apptocmd{\@title}{\par {\large #1 \par}}{}{}
}
\makeatother

\begin{document}
\title{Secure Multi-Party Computation for Inter-Organizational Process Mining}
%
%
\author{Gamal Elkoumy\inst{1} \and
Stephan A. Fahrenkrog-Petersen\inst{2}\and
Marlon Dumas\inst{1} \and\\ Peeter Laud\inst{3} \and Alisa Pankova\inst{3} \and 
Matthias Weidlich\inst{2}}
\authorrunning{G. Elkoumy et al.}
%
\institute{University of Tartu, Tartu, Estonia \\
	\email{\{gamal.elkoumy,marlon.dumas\}@ut.ee} \and
Humboldt-Universität zu Berlin,  Berlin, Germany
\email{\{fahrenks,weidlima\}@hu-berlin.de} \and 
Cybernetica, Tartu, Estonia\\
\email{\{peeter.laud,alisa.pankova\}@cyber.ee}
}

\maketitle              
\begin{abstract}
Process mining is a family of techniques for analyzing business processes based 
on event logs extracted from information systems. Mainstream process mining 
tools are designed for intra-organizational settings, insofar as they assume 
that an event log is available for processing as a whole. The use of such tools 
for inter-organizational process analysis is hampered by the fact that such 
processes involve independent parties who are unwilling to, or sometimes 
legally prevented from, sharing detailed event logs with each other. In this 
setting, this paper proposes an approach for constructing and querying a common 
artifact used for process mining, namely the frequency and 
time-annotated 
Directly-Follows Graph~(DFG), over multiple event logs belonging to different 
parties, in such a way that the parties do not share the event logs with each 
other. The proposal leverages an existing platform for 
secure multi-party computation, namely Sharemind. 
Since a direct 
implementation of DFG construction in Sharemind suffers from 
scalability issues, we propose to rely on vectorization of 
event logs and to employ a divide-and-conquer scheme for parallel 
processing of sub-logs. 
The paper reports on experiments that evaluate the 
scalability of the approach on real-life logs.


\end{abstract}

\section{Introduction}

\begin{figure}[t!]
	\centering
	\includegraphics[width=0.98\textwidth]{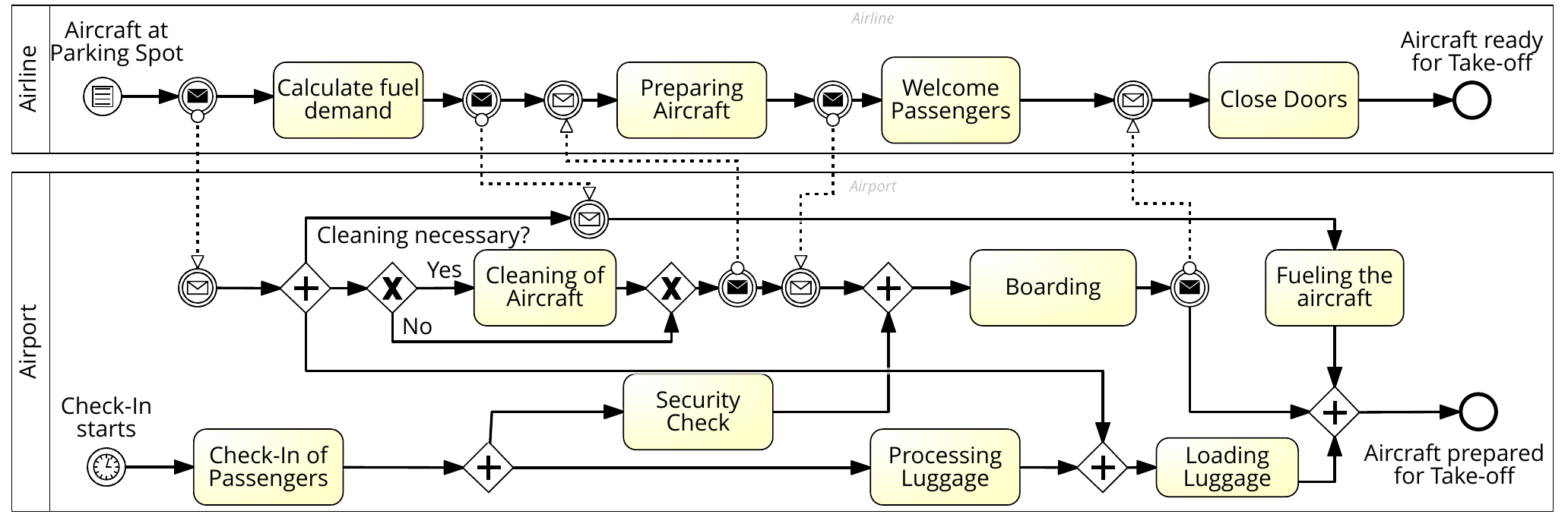}
	\caption{Aircraft ground handling process.}
	\label{fig:ground_handling_process}	
	\vspace{-.6em}
\end{figure}

Contemporary process mining techniques enable users to analyze business processes based on event logs extracted from information systems. The outputs of process mining techniques can be used, for example, to identify performance bottlenecks, waste, or compliance violations. 
Existing process mining techniques require access to the entire event log of a business process.
Usually, this requirement can be fulfilled when the event log is collected from one or multiple systems within the same organization.  
In practice, though, many business processes involve multiple independent organizations. 
We call such processes \emph{inter-organizational business processes}. An example of such process is the process for ground handling of an aircraft, as illustrated in \autoref{fig:ground_handling_process}. This process involves two parties: the airline and the ground handler (called ``airport'' in the model). 

Due to confidentiality concerns as well as privacy regulations, such as GDPR\footnote{\url{https://eur-lex.europa.eu/eli/reg/2016/679/oj}} and HIPAA\footnote{\url{https://www.hhs.gov/hipaa/}}, it is not always possible for organizations to share process execution data with each other. 
Exchanging execution data may reveal personal information of customers or it may expose business secrets. As a result, common techniques for process mining cannot be employed for inter-organizational business processes. Yet, analyzing these processes is often crucial for improving operational performance. With reference to the above scenario, the  effective coordination of ground handling activities is crucial for both involved parties. It determines the number of flights an airport can operate and the number of flights an airline can offer. At the same time, each of the parties needs to protect their confidential data.



In this paper, we focus on the question of how to enable process mining for 
inter-organizational business processes without requiring the involved parties 
to share their private event logs or trust a third party.
To this end, we propose an architecture for process mining based 
on \emph{secure multi-party computation}~(MPC)~\cite{yao1982protocols}. In essence, MPC aims 
at the realization of some computation over data from multiple parties, while
exposing only the result of the computation, but keeping the input data 
private. 
We consider the setting of an MPC platform where the 
involved parties upload their event logs to a network of compute nodes. 
Before the upload, secret sharing algorithms locally 
split each single data value into different parts (i.e., shares) that are then 
stored at different nodes. Since each share does not provide any 
information about the original data, the uploaded event log is encrypted and 
exposed neither to the platform operator nor other involved parties. 
Nonetheless, the MPC platform enables the computation over the encrypted data 
through protocols for result sharing among the nodes. 

We realize the above architecture to answer analysis queries that are common in 
process mining. Specifically, we show how to construct a frequency and time-annotated 
Directly-Follows Graph (DFG), which is a starting point for process discovery 
algorithms  and  performance analysis. While keeping the computed DFG private, we are revealing only the output of performance analysis queries such as finding the top-k bottlenecks (i.e.\ activities with longer cycle time) or the top-k most frequent hand-offs.
We implement our proposed architecture using the Sharemind 
platform~\cite{bogdanov2008sharemind}. In order to tackle scalability issues 
that would be imposed by a naive implementation, we employ vectorization of 
event logs and propose a divide-and-conquer scheme for parallel 
processing of sub-logs. We test the effectiveness of these optimizations 
in experiments with real-world event logs. 

The remainder of the paper is structured as follows. \autoref{sec:background} 
lays out 
related work and the background for this work. \autoref{sec:approach} 
introduces our architecture for privacy-preserving inter-organizational process 
mining along with the optimizations needed for efficient implementation. An 
experimental evaluation is presented in \autoref{sec:evaluation}, before 
\autoref{sec:conclusion} concludes the paper.

\section{Background and Related Work}
\label{sec:background}
In this section, we review work on privacy-preserving process mining, 
inter-organizational process mining, and secure multi-party computation. 
\subsection{Privacy-Preserving Process Mining}
\label{sec:privacy_process_mining}

The necessity of privacy-preserving process mining, due to legal 
developments such as the GDPR, was recently discussed 
in~\cite{mannhardt2018privacy,pika2019towards}. In general, two 
approaches have been established~\cite{caise/Fahrenkrog-Petersen19}: 
(i) anonymizing the event data to apply standard techniques to it, and (ii) 
directly incorporating privacy considerations in process mining 
techniques. The anonymization of event logs from one organization may be done 
using algorithms, like \emph{PRETSA}~\cite{icpm/Fahrenkrog-Petersen19},  that provide privacy guarantees such as
$k$-anonymity~\cite{sweeney2002k}. 
These notions, based on data similarity, 
are widely adopted and offer protection against certain 
attacks like the 
disclosure of the identity of individuals involved in the dataset.  
Approaches based on cryptography~\cite{burattin2015toward,simpda/RafieiWA18} 
have also been proposed. Following the idea to incorporate privacy guarantees 
directly in process mining techniques, algorithms for privacy-preserving 
process discovery~\cite{MannhardtKBWM19,tillem2017mining} have been 
proposed. 
Recently, techniques for privacy-preserving process mining, following either of 
the aforementioned paradigms, have 
been made available for a large audience with the tool 
ELPaaS~\cite{elpaas2019}. 

All of the above techniques, with the exception 
of~\cite{tillem2017mining}, concern process mining for a single organization 
and have not yet been adopted or evaluated for an 
inter-organizational setting. While the focus of \cite{tillem2017mining} is on 
an inter-organizational setting, the approach targets solely the creation of 
a process model, while we aim at answering a wide range of analysis queries 
about business processes.

\subsection{Inter-Organizational Process Mining}
\label{sec:cross_orgranizational}

The problem of automated discovery of process models in an inter-organizational setting has been considered in~\cite{schulz2004facilitating,zeng2013cross}. However, these approaches do not address privacy concerns. 
Similarly, another line of related research proposes techniques to compare executions of the same process across multiple organizations~\cite{aksu2016cross,buijs2011towards}, but without considering privacy requirements.

The problem of ensuring privacy in inter-organizational process mining has been addressed by Liu et al.~\cite{Liu19}. They provide a process mining framework based on the assumption that the parties in the process are willing to share confidential information with a third (trusted) party. This assumption is unrealistic in many situations. In this paper, we address the problem of inter-organizational process mining in the context where the parties in the process are unwilling to share any execution data with each other or with a third party. In one of the embodiments of our proposal, a third party is involved for computation purposes, but this third party does not get access to any information during this computation.


 \subsection{Secure Multi-Party Computation}
 \label{sec:mpc}

 Secure Multi-party Computation (MPC)~\cite{GMW} is a cryptographic functionality that allows $n$ parties to cooperatively evaluate $(y_1,\ldots,y_n)=f(x_1,\ldots,x_n)$ for some function $f$, with the $i$-th party contributing the input $x_i$ and learning the output $y_i$, and no party or an allowed coalition of parties learning nothing besides their own inputs and outputs.
There exist a few approaches for constructing MPC protocols.
%
%
Homomorphic secret sharing~\cite{shamir1979share} is a common basis for MPC protocols. In such protocols, the arithmetic or Boolean circuit representing $f$ is evaluated gate-by-gate, constructing secret-shared outputs of gates from their secret-shared inputs. Each evaluation requires some communication between parties (except for addition gates), hence the depth of the circuit determines the round complexity of the protocol. On the other hand, there exist protocols with low communication complexity~\cite{DBLP:conf/ccs/ArakiFLNO16}, allowing the secure computation of quite complex functions $f$, as long as the circuit implementing it has a low multiplicative depth.

The complexity of MPC protocols is dependent on the number of parties 
jointly performing the computations. Hence the typical deployment of MPC has a 
small number of compute nodes, also known as \emph{computation 
parties}, which execute the protocols for evaluating 
gates, while an unbounded number of parties may contribute the inputs and/or 
receive the outputs of the computation. Several 
frameworks support such deployments of MPC and provide APIs to simplify the 
development of privacy-preserving applications~\cite{archer2018keys}. One of 
such frameworks is Sharemind~\cite{bogdanov2008sharemind}, whose main protocol 
set is based on secret-sharing among 
three computing parties. In this paper, we build on top of Sharemind, 
 but our techniques are also 
applicable to other secret sharing-based MPC systems. 

In Sharemind, a party can play different roles: an input party, a computation party, and/or an output party. In the case where only two parties are involved in an inter-organizational process, these two parties play the role of input parties and also that of computing parties. To fulfill the requirements of Sharemind, they need to enroll a third computing node, which merely performs computations using secret shares from which it can infer no information.\footnote{When three or more parties are involved in a  process, no external party is required.} 

The Sharemind framework provides its own programming language, namely the SecreC language~\cite{bogdanov2014domain}, for programming privacy-preserving applications. SecreC allows us to abstract away certain 
details of cryptographic protocols.

\newcommand{\share}[1]{[\![{#1}]\!]}

\section{Multi-Party Computation based Process Mining}
\label{sec:approach}
This section introduces our techniques for process mining based on 
secure multi-party computation. \autoref{sec:model} first 
clarifies our model for inter-organizational process mining including the 
required input data and the obtained analysis results. We then introduce our 
architecture for realizing the respective analysis using secure multi-party 
computation in \autoref{sec:core_approach}. In 
\autoref{sec:div_and_conq}, we elaborate on vectorization and 
parallelization to improve the efficiency of our approach.

\subsection{Model for Inter-organizational Process Mining}
\label{sec:model}

We consider a model in which an event log $L=\{e_1,\ldots,e_n\}$ is defined as 
a set of events $e=(i,a,ts)$, each capturing a single execution of an activity 
$a$ at time $ts$, as part of a single instance $i$ of the business process. 
Grouping events by the latter and ordering them according to their timestamp 
enables the construction of traces $t=\langle 
e_1,\ldots,e_m \rangle$, i.e., single executions of the process, so that we 
refer to $i$ also as the trace identifier. 


For an inter-organizational business process, an event log that records 
the process execution from start to end is commonly not available. Rather, 
different parties record sub-logs, built of events that 
denote activity executions at the respective party. To keep the notation 
concise, we consider a setting in which two parties, $I_a$ and $I_b$, execute 
an 
inter-organizational process, e.g., the airport and the airline in our 
motivating example. Then, each of the two parties records an event log, denoted 
by $L_a$ and $L_b$. 
Each of these logs is the projection of $L$ on the events that 
denote activity executions at the respective parties $I_a$ and $I_b$. We assume 
that each activity can only be executed by one of the parties, so that this 
projection is defined unambiguously. 

For the above setting, we consider the scenario that the parties $I_a$ and 
$I_b$ want to answer some analysis queries $Q$ over the inter-organizational 
event log $L$, yet \emph{without} sharing their logs $L_a$ and $L_b$ with each 
other. 
More specifically, we focus on analysis queries that can be answered on a 
frequency or time-annotated DFG of the inter-organizational 
process. The basic DFG captures the frequencies with which the executions 
of two activities have been observed to directly follow each other in a 
trace. Moreover, we consider temporal annotations of the directly-follows 
dependencies in terms of time between the respective activity executions. 
Queries over the frequency and time-annotated DFGs allow us to analyze the main 
paths of the process, the rarely executed paths, as well as the activities that most contribute to delays in a process. Note though that only query answers are to be revealed whereas the 
actual DFG shall be kept private. 

Formally, the time-annotated DFG is captured by an $|A| \times |A|$ matrix, 
where $A$ is the set of all possible activities of the process. Each cell 
contains a tuple $(c,\Delta)$. The counter $c$ represents the frequency with 
which a directly-follows dependency has been observed in $L$, i.e., for the 
cell $(a_1,a_2)$ it is the number of times that two events $e_1=(i_1,a_1,ts_1)$ 
and $e_2=(i_2,a_2,ts_2)$ follow each other directly in some trace (i.e., $i_1 = 
i_2$) of $L$. Also, 
$\Delta$ is the total sum of the time passed by between all occurrences of 
the respective events, i.e., $ts_2 - ts_1$ for the above events. 

In inter-organizational process mining, the above time-annotated DFG 
cannot be computed directly, as this would require the parties to share their sub-logs.

\subsection{MPC Architecture for Process Mining}
\label{sec:core_approach}

\begin{figure}[!htb]
	\centering
	\includegraphics[width=.98\columnwidth]{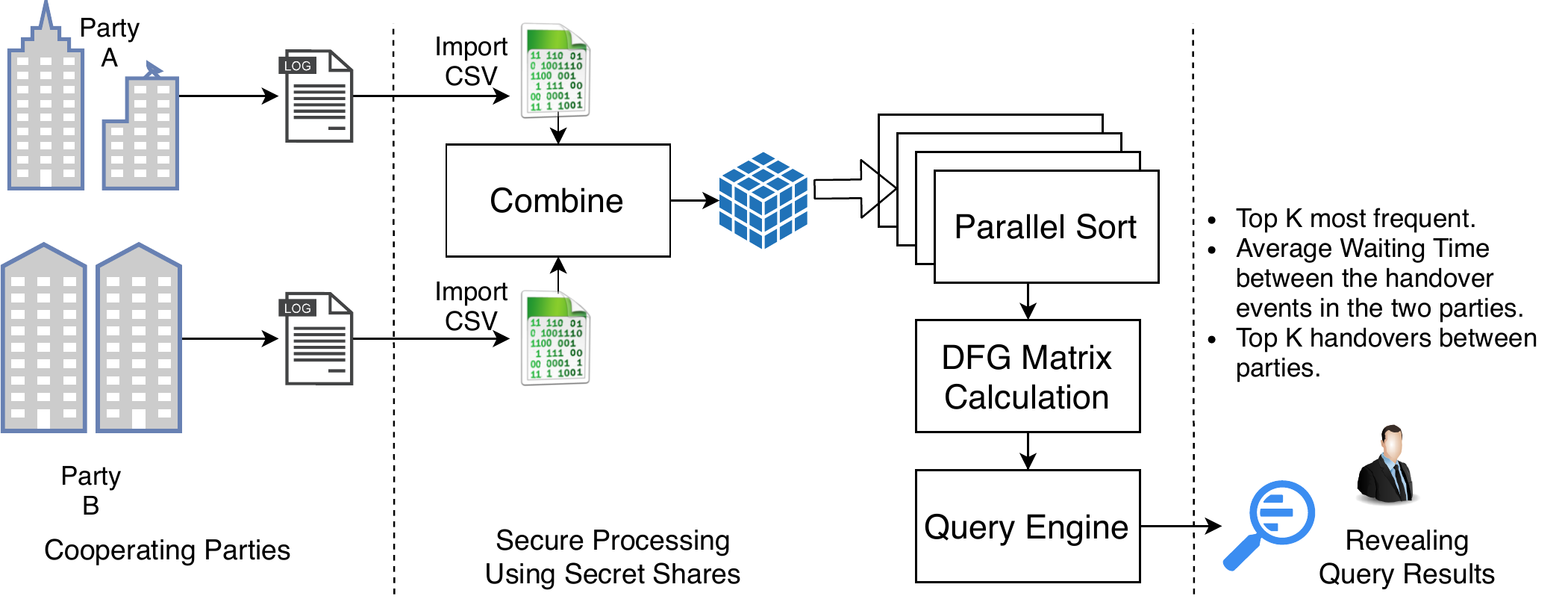}
	\caption{Overview of the proposed approach}
	\label{fig:overview}
\end{figure}

\begin{figure*}[t!]
	\centering
	\begin{subfigure}[b]{0.9\textwidth}
		\centering
		\includegraphics[width=\textwidth]{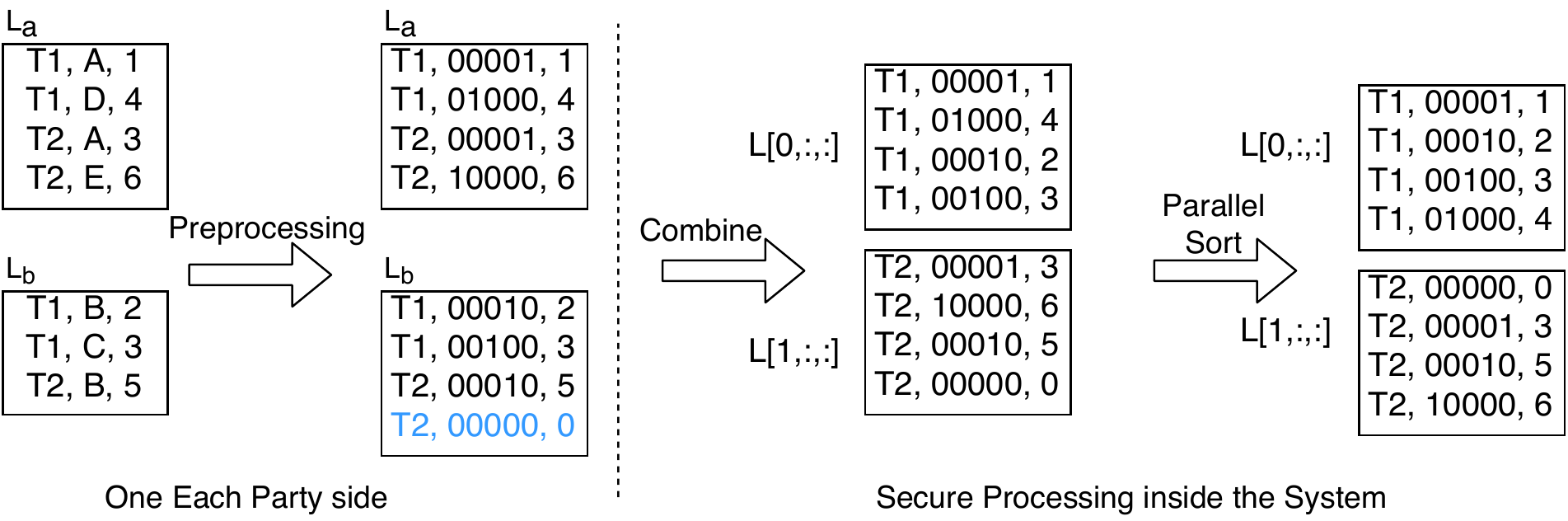}
		\caption{Illustration of the preprocessing, combine and parallel sort 
		steps}
		\label{fig:example_input}
	\end{subfigure}

	\begin{subfigure}[b]{0.3\textwidth}
		\centering
		\includegraphics[width=\textwidth]{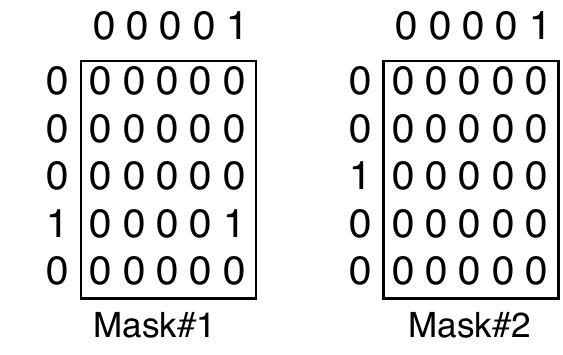}
		\caption{Example of two masks}
		\label{fig:example_masks}
	\end{subfigure}
	~
	\begin{subfigure}[b]{0.38\textwidth}
	\centering
	\includegraphics[width=\textwidth]{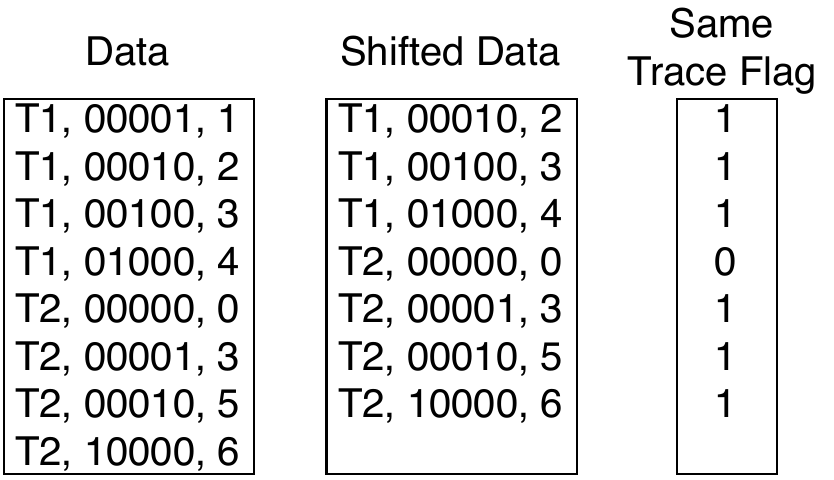}
	\caption{DFG Calculation using shift and a flag}
	\label{fig:example_shifted}
\end{subfigure}
	~
	\begin{subfigure}[b]{0.22\textwidth}
		\centering
		\includegraphics[width=\textwidth]{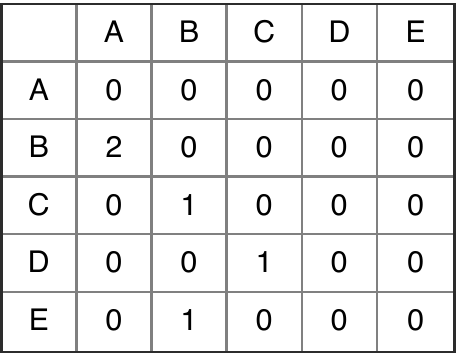}
		\caption{The DFG matrix with counts}
		\label{fig:example_dfg}
	\end{subfigure}
	\caption{Example of two event logs and their processing steps inside the system}
	\label{fig:running_example}
\end{figure*}

To enable inter-organizational process mining \emph{without} requiring parties 
to share their event logs 
with each other, we propose an architecture based on secure multi-party 
computation (MPC). As outlined in \autoref{fig:overview}, we rely on a platform 
for MPC (in our case Sharemind~\cite{bogdanov2008sharemind})
that takes the event logs of the participating parties, i.e., $L_a$ and $L_b$, 
as secret-shared input. Inside the MPC platform, the respective data is processed 
in a privacy-preserving way in order to answer analysis queries over the time-annotated 
DFG computed from that data. In  \autoref{fig:running_example}, we present a running example of the processing steps of the system.


Below we summarize the functionality enbodied in the proposed MPC platform for inter-organizational process mining.

\paragraph{Preprocessing.} Each party performs the preparation of its log 
	at its own site. The parties share the number of unique activities and the 
	maximum number of events per trace. In \autoref{fig:example_input}, we show 
	an example with two traces.  In the preprocessing step, all traces are 
	padded to the same length, as illustrated with the blue event in 
	\autoref{fig:example_input}. The activities are transformed into a one-hot 
	encoding that is used for masking at the DFG calculation step, as will be 
	explained later. The logs are 
	sorted by traces.
	
\paragraph{Combination.} The parties upload their event logs 
$L_a$ and $L_b$ to the MPC platform in a secret-shared manner. That is, the 
values $(i,a,ts)$ of each event (encoded as integers) are split into shares, 
which do not provide any information on the original values and are stored at 
different nodes of the platform. This way, each party can only see the total 
number of records uploaded by each party, but not the particular data. 
Subsequently, the logs are unified, creating a single log of events ${L}$. The 
combination is performed in a manner to divide the logs into processing chunks. 
As long as we are making the number of events per trace is fixed, that is 
possible by dividing the index by the number of traces for each event and 
assigning data from the same trace to the same chunk. In 
\autoref{fig:example_input}, the system processes one trace with its own 
chunk.  
 
\paragraph{Sorting.} To calculate the annotated DFG, we have to determine 
which events follow each other in a trace by grouping the events by their 
trace identifier and ordering them by their timestamp. Since the trace 
identifier is secret-shared, we cannot group events directly. Instead, we use a
privacy-preserving quicksort algorithm~\cite{Hamada12} 
as implemented in Sharemind to sort the events by 
their trace identifier. Applying the same algorithm also to the secret-shared 
timestamps ensures that the events of the same trace follow each other in 
the order of their timestamps, which is illustrated as the last step in 
\autoref{fig:example_input}.

\paragraph{DFG matrix calculation.} Next, we construct the DFG matrix inside 
the MPC platform, keeping it secret. Since the information on the activity of 
an event is secret-shared, we cannot simply process the events of traces 
sequentially as the matrix cell to update would not be known. 
Hence, we adopt a one-hot encoding for activities, so that each possible 
activity is represented by a binary vector of length $|A|$. To mask the actual 
number of possible activities, the set over which the vector is defined may 
further include some padding, i.e., the vector length can be chosen to be 
larger than $|A|$. 
Now, if we compute the \emph{outer product} of such vectors for activities 
$a_1$ and $a_2$, we get a mask matrix $M$ such that $M[a_1,a_2] = 1$, while all 
other entries are $0$. An example of such masks is given in 
\autoref{fig:example_masks}. The first mask represents the 
directly-follows dependency from activity $A$ to $B$ of our running example. 
The 
second mask encodes the directly-follows dependency from activity $A$ to $C$.  
For all sequential pairs of events in the sorted log, we 
sum up these matrices to get the frequency count $c$ of the 
directly-follows dependency for $(a_1,a_2)$. 
Multiplying $M$ by the duration between two events further enables 
us to derive the total sum duration passed, i.e., $\Delta$, of the 
directly-follows-relation. The duration operation is performed between every two consecutive events of the same trace. We can perform the duration calculation by using an element-wise vector subtraction by duplicating the 
dataset and then shifting its events by one as in 
\autoref{fig:example_shifted} 
. Technically, the outer product is a 
 function 
that is realized as a protocol over secret-shared data in Sharemind, and its 
runtime complexity is linear in $|A|$~\cite{laud2017privacy}.

However, the above approach could mix up events of different traces. We 
therefore also compute a flag $b$ that is $1$, if the trace identifiers of two 
events are equivalent, and $0$ otherwise, which is illustrated as the "Same 
Trace Flag" column in \autoref{fig:example_shifted}. Then, we multiply the mask 
matrix $M$ 
by $b$, so that the values of $M$ are ignored, if $b = 0$. Again, the  
functionality for comparison and multiplication can be traced back to 
predefined protocols in 
Sharemind. We show the DFG matrix with counts of our running example in 
\autoref{fig:example_dfg}.

\autoref{alg:dfg} summarizes the computation of the annotated DFG from the 
sorted, combined log $L$, where $\share{\cdot}$ denotes a secret-shared data 
value.

\paragraph{Query answering.} A query $Q$ defines a subset $S$ of the 
annotated DFG, which is generated by the MPC platform and revealed to the 
participating parties. Through sharing the $S$ solely, but not the complete 
annotated DFG, we are able to limit the amount of information each party can 
learn about the process. As an example, consider the query to derive the average waiting time between the handover events between the two parties. Based on the secret-shared DFG, the respective 
activities may be identified through grouping and sorting the events, similar 
to the procedure outlined above, which is again based on the predefined 
protocols of an MPC platform such as Sharemind. 



\begin{algorithm}[t!]
	\caption{Calculating the combined, annotated DFG($\share{L}$)}\label{alg:dfg}
	\label{alg:merge}
	\begin{flushleft}
		\textbf{INPUT:} The sorted, combined event log $\share{L}$ of length 
		$n$.\\
		\textbf{OUTPUT:} Annotated DFG comprising a count matrix $\share{G}$ 
		and a time matrix $\share{W}$.
		\vspace{-.4em}
	\end{flushleft}
	\begin{algorithmic}[1]
		\State Initialize $\share{G} = 0$, $\share{W} = 0$
		\ForAll{$j \in \{1,\ldots,n-1\}$} \label{line:all_cases}
		\State $\share{b}\leftarrow (\share{L[j-1].i} = \share{L[j].i})$; 
		\hfill//compute the flag for traces
		\State $\share{M}\leftarrow \share{b} \cdot \left(\share{L[j-1].a} 
		\otimes 
		\share{L[j].a}\right)$; \hfill//compute the outer product
		\State $\share{G}\leftarrow\share{G} + \share{M}$; \hfill//incorporate 
		the current dependency
		\State $\share{W}\leftarrow\share{W} + \share{M} \cdot (\share{L[j].ts} 
		- \share{L[j-1].ts})$; \hfill//Incorporate the time lag
		\EndFor
		\State \Return $\share{G}$, $\share{W}$
	\end{algorithmic}
\end{algorithm}



\subsection{Performance Optimizations}
\label{sec:div_and_conq}

Inter-organizational process mining using the above general architecture might 
suffer from scalability issues. The reason is that privacy-preserving  
computation through protocols over secret-shared data is inevitably less 
efficient than plain computation. Hence, even 
for functions that have a generally low runtime complexity ($\mathcal{O}(n)$ 
for the combination, $\mathcal{O}(n\log(n))$ for the sorting, 
$\mathcal{O}(nm^2)$ for the calculation of the annotated DFG, where $n$ is the 
log length and $m$ is the number of activities), there is a non-negligible 
overhead induced by MPC. For instance, a naive realization of the quicksort 
algorithm to sort events would require 
$\mathcal{O}(n\log(n))$ rounds of communication between the nodes and 
$\mathcal{O}(n\log(n))$ value comparisons per round~\cite{Hamada12}. We 
therefore consider two angles to improve the efficiency of 
the analysis, namely vectorization and parallelization. 


\paragraph{Vectorization.} A computation that adopts a single-instruction 
multiple-data (SIMD) approach is highly recommended in MPC applications. Since 
MPC assumes continuous interaction between distributed nodes, the number of 
communication rounds shall be reduced as much as possible. For instance, while 
computing $n$ multiplications sequentially would result in $n$ rounds of 
communication, one may alternatively multiply element-wise two vectors of 
length $n$, for which one round of network communication is sufficient. 
Sharemind offers efficient protocols for such vector-based 
functions~\cite{laud2017privacy}.



\paragraph{Parallelization.}
Further runtime improvements are obtained by parallelizing the algorithm 
itself. Again, our goal is to reduce the number of rounds of communication 
among the nodes of the MPC platform. We, therefore, split the input data into 
\emph{chunks}, such that all chunks can be processed independently from each 
other. In our scenario, this is done by grouping the party logs by trace, or by 
a group of traces, generating an annotated DFG per group, and finally 
integrating the different DFGs. 
Since events of the same trace will never occur in different chunks, instead 
of sorting one log of length $n$, we will need to sort $c$ chunks of length 
$n/c$ each. Since the communication complexity of a privacy-preserving quicksort is 
$O(n\cdot \log{n})$~\cite{Hamada12}, this improves efficiency.

The above approach raises the question of determining the size of the chunks. 
Separating each trace reveals the total number of events of that trace provided 
by a party, which may be critical from a privacy perspective. On the other 
hand, a small chunk size reduces the overhead of sorting. This leads to a 
trade-off between runtime performance and privacy considerations.

However, in our current implementation, all chunks must have 
the same length, as Sharemind allows parallel sorting only for equal-length 
vectors. Therefore, we apply padding to the traces in the log, adding dummy 
events (for which an empty vector in the one-hot encoding represents the activity so that the events are ignored for the DFG calculation) until the 
number of events of the longest trace is reached. Such padding may be employed 
locally, by each party, and also has the benefit that the length of individual 
traces is not revealed. 


\section{Evaluation}
\label{sec:evaluation}

We implemented the proposed approach on top of the Sharemind multi-party computation platform.\footnote{\url{https://sharemind-sdk.github.io}} The source code of our implementation is available at \url{https://github.com/Elkoumy/shareprom}. The implementation is written using the SecreC programming language supported by Sharemind. 

Using this implementation, we conducted feasibility and scalability experiments, specifically to address the following research questions:
 \begin{description}
 	\item[RQ1:] How do the characteristics of the input event logs influence the performance of the secure multi-party computation of the DFG?
 \item[RQ2:] What is the effect of increasing the number of parallel chunks on the performance of the multi-party computation of the DFG?
\end{description}

\subsection{Datasets}
\label{sec:data}

The proposed approach is designed to compute the DFG of an inter-organizational process where the event log is distributed across multiple parties, and each party is responsible for executing a subset of the activities (i.e.\ \emph{event types}) of the process. We are not aware of publicly available real-life datasets with this property. 
We identified a collection of synthetic inter-organizational business process event logs~\cite{borkowski2017event}. However, these logs are too small to allow us to conduct performance experiments (a few dozen traces per log). On the other hand, there is a collection of real-life event logs of intra-organizational processes comprising logs of varying sizes and characteristics \footnote{\url{https://data.4tu.nl/repository/collection:event_logs_real}}. From this collection, we selected three logs with different size and complexity (cf.~\autoref{tab:event_logs}):
\begin{description}
	\item[BPIC 2013] This event log captures incident and problem management process at an IT department of a car production company.
		\item[Credit Requirement] This event log comes from a process for background checking for the purpose of granting credit at a Dutch bank. 
		 It has a simple control-flow structure: All traces follow the same sequence of activities.
	\item[Traffic Fines] This event log comes from a process to collect payment of fines from traffic law violations at a local police office in 
	Italy. 
\end{description}

\begin{table}[hbtp]
	\centering
	\caption{Event Logs for Evaluation}
	\label{tab:event_logs}
	\footnotesize
	\begin{tabular}
		{lr@{\hspace{.7em}}r@{\hspace{.7em}}r@{\hspace{.7em}}r@{\hspace{.7em}}
			r@{\hspace{.7em}}r}
		\toprule
		Event Log & \# Events & \# Cases & \# Activities & 
		\multicolumn{3}{c}{\# Events in Case}\\
		\cmidrule{5-7}
		& & & & Avg & Max & Min \\
		\midrule
		BPIC 2013 &	6,660 &	 1,432 &6  & 4.478& 35 & 1\\
		Credit Requirement &  50,525 &	 10,034 &8 & 15 &15 & 15\\
		Traffic Fines &	 561,470 & 	150,370 &  11 &3.73& 20 & 2\\
		\bottomrule
	\end{tabular}
\end{table}

To simulate an inter-organizational setting, we use a round-robin approach to 
assign each event type (activity) in the log to one of two parties. Hence, each 
party executes half of the event types.

\subsection{Experimental Setup}
\label{sec:exp_setup}

To answer the above questions, we use the following performance measures:
\begin{itemize}
\item 
\textbf{Runtime.} We define runtime as the amount of time needed to transform 
the event logs of the two parties securely into an annotated DFG. 
 We also report the throughput, the number of events processed by the system per second, to provide a complementary perspective.
\item 
\textbf{Communication Overhead.} 
We define the communication overhead as the amount of data transferred between 
the computing parties during the multi-party computation. We measure this overhead as the volume of the data sent and received. 
The communication overhead gives insights into how much the performance of the multi-party computation would degrade if the computing nodes of the parties were distributed across a wide-area network
.
\end{itemize}

We performed five runs per dataset per experiment.
 We report the average maximum 
values for latency and the average value for both throughput and communication overhead, across the five runs. 
We used \emph{Nethogs}\footnote{\url{https://github.com/raboof/nethogs}} to 
measure the communication overhead, and we report the average value per compute node. The experiments were run in an environment with three physical servers as compute nodes with Sharemind installed on them. Each server has an AMD Processor 6276 and 192 GB RAM. The servers are connected using a 1GB Ethernet switch.

The experiments focus on the time needed to construct the annotated DFG, since 
it is the most sophisticated and time-consuming portion of the proposed analysis 
pipeline, due to the communication required between the compute nodes. Once the annotated DFG is available, stored in a secret-shared manner, the calculation of the actual queries has a lower complexity.

\subsection{Results}
\label{sec:results}

\begin{figure*}[t!]
	\centering
		\begin{subfigure}[b]{0.90\textwidth}
		\centering
		\includegraphics[width=.95\columnwidth]{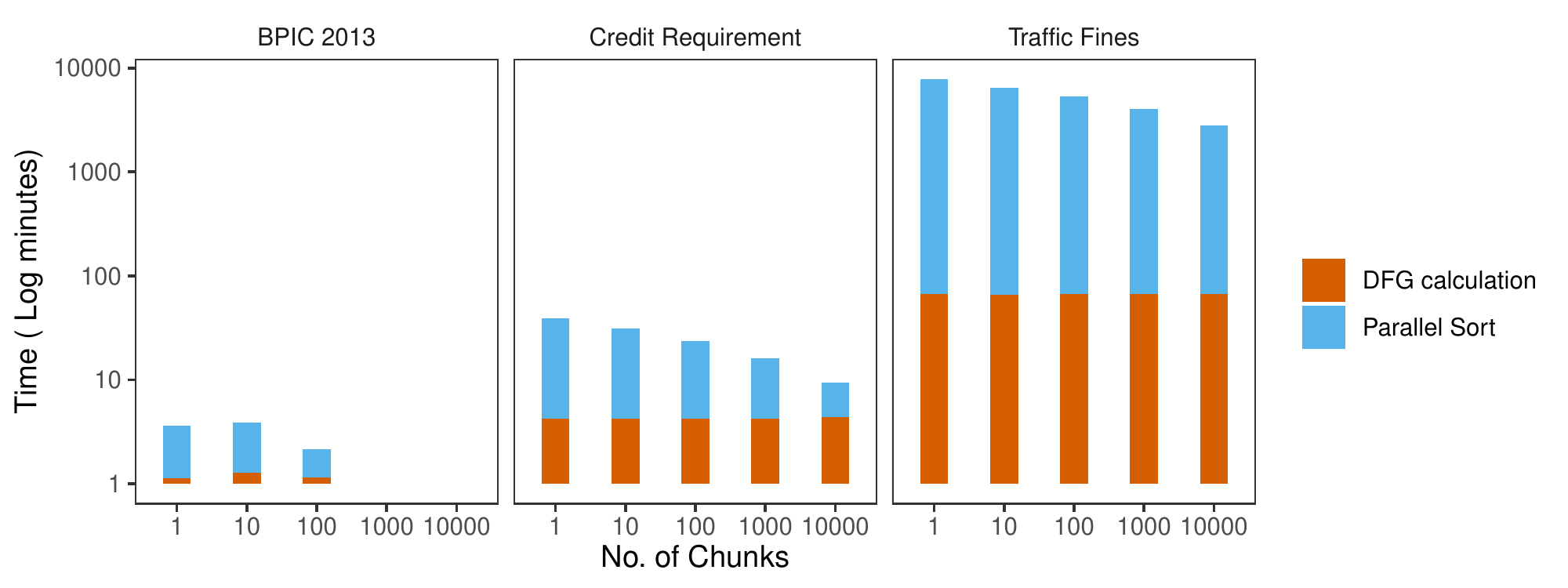}
		\caption{Runtime Experiment: Execution Time (Log) vs no. of Chunks.}
		\label{fig:latency}
	\end{subfigure}
	
	\begin{subfigure}[b]{0.70\textwidth}
		\centering
		\includegraphics[width=.90\columnwidth]{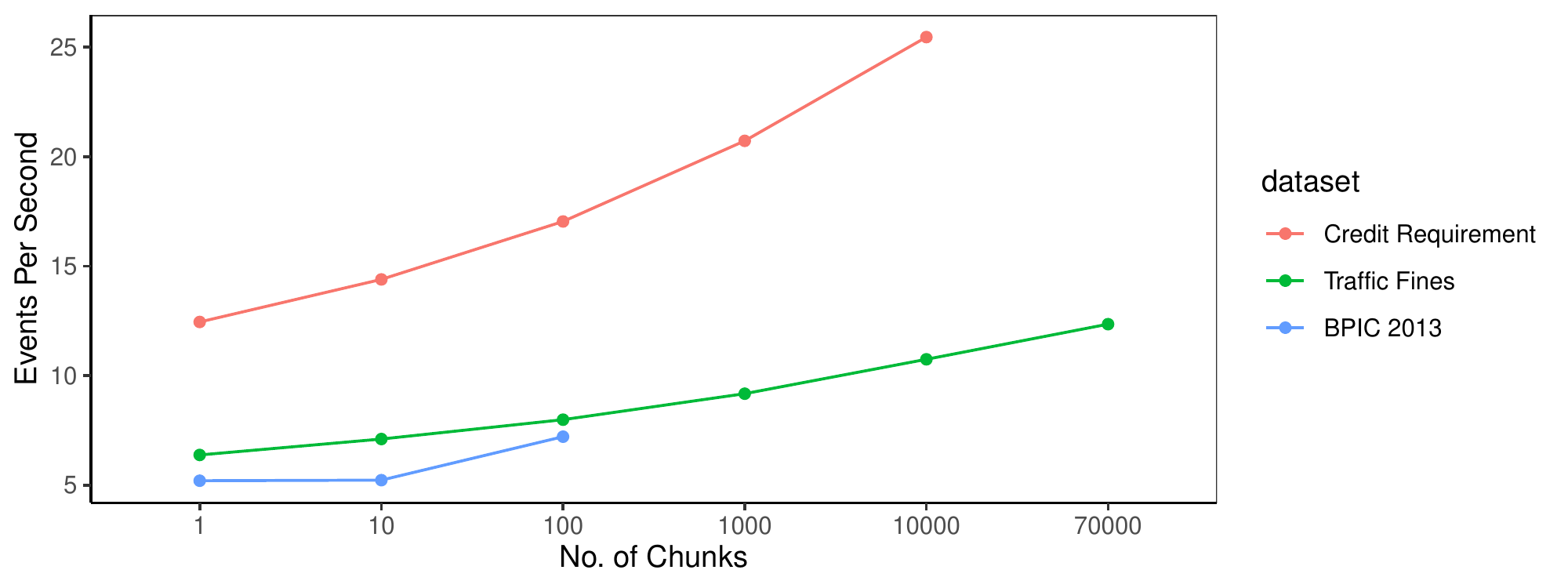}
		\caption{Throughput Experiment: Events per Second vs no. of Chunks.}
		\label{fig:throughput}
	\end{subfigure}

	\begin{subfigure}[b]{0.9\textwidth}
		\centering
		\includegraphics[width=.95\columnwidth]{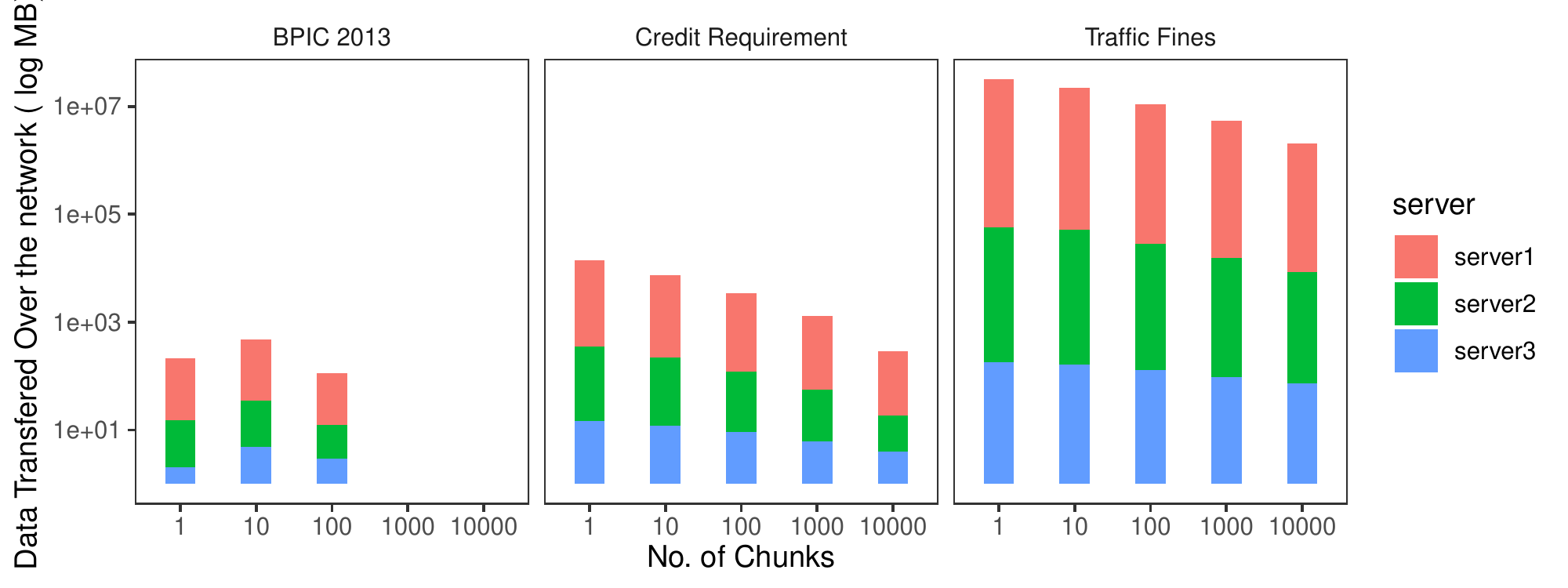}
		\caption{Communication Overhead Experiment: Data Transferred (Log) vs no. of Chunks.}
		\label{fig:comm}
	\end{subfigure}
	\caption{Experimental Evaluation of the Proposed Approach}
	\label{fig:experiment}
\end{figure*}

\paragraph{Runtime Experiment.} In \autoref{fig:latency}, we illustrate the 
observed execution time when varying the number of chunks used in the 
parallelization. We plot a bar for each chunk size. Each bar represents the 
runtime of the parallel sort in blue and the run time of the DFG  
calculation in orange. From \autoref{fig:latency}, we conclude that the 
runtime decreases with an increasing number of chunks, due to 
the parallel sorting of chunks. We also note that the runtime for the DFG calculation stays constant. 
In \autoref{fig:throughput}, we report the 
number of processed  events per second when varying the number of chunks. We 
find a consistent improvement for the throughput across all event logs.

Regarding RQ1, we summarise that the proportion 
of runtime between sorting and DFG calculation differs based on the 
event log characteristics. For the log with the largest number of event types, 
the DFG calculation makes up the most substantial proportion of the total runtime. In contrast, 
the proportion is significantly lower for the 
logs with a smaller number of event types. A possible explanation for this finding is the increasing size of the vectors required to represent each activity due to our bit-vector representation. Such increase results in more computational heavy 
calculations. 
Regarding RQ2, we conclude that the runtime decreases for 
event logs with an increasing number of chunks.


\paragraph{Communication Overhead.} In \autoref{fig:comm}, we present the 
amount of data transferred to each server, again also varying the number of 
chunks. 
We observe that the communication overhead decreases with an increase in 
the number of chunks. These findings confirm our earlier findings regarding 
RQ2. In summary, a higher number of chunks leads to improved performance across all three measures. 

\paragraph{Threats to validity.} 
The evaluation reported above has two limitations. First, the event logs used in the evaluation, while coming from real-life systems, are intra-organizational event logs, which we have split into separate logs to simulate an inter-organizational setting. It is possible that these logs do not capture the communication patterns found in inter-organizational processes. Second, the number of event logs is reduced, which limits the generalizability of the conclusions.
The results suggest that the proposed technique can handle small-to-medium-sized logs, with relatively short traces, but there may be other characteristics of event logs that affect the performance of the proposed approach. 




\section{Conclusion}
\label{sec:conclusion}
This paper introduced a framework for inter-organizational process mining based 
on secure multi-party computation. The framework enables two or more parties to 
perform basic process mining operations over the partial logs of an 
inter-organizational process held by each party, without any information being 
shared besides: (i) the output of the queries that the parties opt to disclose; 
and (ii) three high-level log statistics: the number of traces per log, the 
number of event types, and the maximum trace length. The paper specifically 
focuses on the computation of the DFG, annotated with frequency and temporal information. This is a basic structure 
used by process mining tools to perform various operations, including automated 
process discovery and various performance analysis queries (e.g. top-k 
bottlenecks and least-frequent and most-frequent flow dependencies).

To mitigate the high-performance overhead commonly observed for secure 
multi-party computation, we introduced two optimizations over the 
basic DFG computation algorithm: one based on vectorization of the event log 
and the other based on a divide-and-conquer strategy, where the log is 
processed in chunks.

An evaluation using real world event logs shows that with these optimizations, 
it is possible to compute the DFG of real-life logs with execution times that 
make this technique usable in practice. 
The divide-and-conquer approach provides opportunities to scale up the proposed 
technique by using a map-reduce execution-style, however not to a sufficient 
level to enable interactive process mining (which requires execution times in 
the order of seconds). Also, the approach is not able to handle logs with thousands of traces. 


In future work, we will explore further optimizations to address these limitations, for example, by taking into account metadata about the event types in the event  log where hand-offs occur between participants. Usually, such event types are 
known as they correspond to message exchange. Therefore, it becomes possible to 
split the logs into a ``private'' part and a ``public'' part (the latter being 
the points where hand-offs occur), and to process them separately using 
different approaches. 

Another avenue for future work is to combine the proposed approach with approaches that provide complementary guarantees such as differential privacy techniques. The latter techniques allow us to noisify the DFG or the outputs from the queries of the DFG to limit the information leaked by these outputs.




\medskip\noindent\textbf{Acknowledgments.} This research is partly funded by ERDF via the Estonian Centre of Excellence in ICT (EXCITE) and the IT Academy programme.

 \bibliographystyle{splncs04}
 \bibliography{bib_mpc_process_discovery}

\begin{thebibliography}{10}
\providecommand{\url}[1]{\texttt{#1}}
\providecommand{\urlprefix}{URL }
\providecommand{\doi}[1]{https://doi.org/#1}

\bibitem{Aalst16}
van~der Aalst, W.M.P.: Process Mining - Data Science in Action, Second Edition.
  Springer (2016)

\bibitem{aksu2016cross}
Aksu, {\"U}., Schunselaar, D.M., Reijers, H.A.: A cross-organizational process
  mining framework for obtaining insights from software products: Accurate
  comparison challenges. In: 2016 IEEE 18th Conference on Business Informatics
  (CBI). vol.~1, pp. 153--162. IEEE (2016)

\bibitem{DBLP:conf/ccs/ArakiFLNO16}
Araki, T., Furukawa, J., Lindell, Y., Nof, A., Ohara, K.: High-throughput
  semi-honest secure three-party computation with an honest majority. In:
  Proceedings of the 2016 {ACM} {SIGSAC} Conference on Computer and
  Communications Security, Vienna, Austria, October 24-28, 2016. pp. 805--817
  (2016)

\bibitem{archer2018keys}
Archer, D.W., Bogdanov, D., Lindell, Y., Kamm, L., Nielsen, K., Pagter, J.I.,
  Smart, N.P., Wright, R.N.: From keys to databases—real-world applications
  of secure multi-party computation. The Computer Journal  \textbf{61}(12),
  1749--1771 (2018)

\bibitem{elpaas2019}
Bauer, M., Fahrenkrog{-}Petersen, S.A., Koschmider, A., Mannhardt, F., van~der
  Aa, H., Weidlich, M.: Elpaas: Event log privacy as a service. In: Proceedings
  of the Dissertation Award, Doctoral Consortium, and Demonstration Track at
  {BPM} 2019 co-located with 17th International Conference on Business Process
  Management, {BPM} 2019, Vienna, Austria, September 1-6, 2019. pp. 159--163
  (2019)

\bibitem{bogdanov2014domain}
Bogdanov, D., Laud, P., Randmets, J.: Domain-polymorphic programming of
  privacy-preserving applications. In: Proceedings of the Ninth Workshop on
  Programming Languages and Analysis for Security. p.~53. ACM (2014)

\bibitem{bogdanov2008sharemind}
Bogdanov, D., Laur, S., Willemson, J.: Sharemind: A framework for fast
  privacy-preserving computations. In: European Symposium on Research in
  Computer Security. pp. 192--206. Springer (2008)

\bibitem{borkowski2017event}
Borkowski, M., Fdhila, W., Nardelli, M., Rinderle-Ma, S., Schulte, S.:
  Event-based failure prediction in distributed business processes. Information
  Systems  (2017)

\bibitem{buijs2011towards}
Buijs, J.C., van Dongen, B.F., van~der Aalst, W.M.: Towards
  cross-organizational process mining in collections of process models and
  their executions. In: International Conference on Business Process
  Management. pp. 2--13. Springer (2011)

\bibitem{burattin2015toward}
Burattin, A., Conti, M., Turato, D.: Toward an anonymous process mining. In:
  2015 3rd International Conference on Future Internet of Things and Cloud. pp.
  58--63. IEEE (2015)

\bibitem{caise/Fahrenkrog-Petersen19}
Fahrenkrog{-}Petersen, S.A.: Providing privacy guarantees in process mining.
  In: (CAiSE Doctoral Consortium 2019), Rome, Italy, June 3-7, 2019. pp. 23--30
  (2019)

\bibitem{icpm/Fahrenkrog-Petersen19}
Fahrenkrog{-}Petersen, S.A., van~der Aa, H., Weidlich, M.: {PRETSA:} event log
  sanitization for privacy-aware process discovery. In: International
  Conference on Process Mining, {ICPM} 2019, Aachen, Germany, June 24-26, 2019.
  pp.~1--8 (2019)

\bibitem{GMW}
Goldreich, O., Micali, S., Wigderson, A.: How to play any mental game or {A}
  completeness theorem for protocols with honest majority. In: Aho, A.V. (ed.)
  Proceedings of the 19th Annual {ACM} Symposium on Theory of Computing, 1987,
  New York, New York, {USA}. pp. 218--229. {ACM} (1987)

\bibitem{Hamada12}
Hamada, K., Kikuchi, R., Ikarashi, D., Chida, K., Takahashi, K.: Practically
  efficient multi-party sorting protocols from comparison sort algorithms. In:
  Information Security and Cryptology - {ICISC} 2012 - 15th International
  Conference, Seoul, Korea, November 28-30, 2012, Revised Selected Papers. pp.
  202--216 (2012)

\bibitem{laud2017privacy}
Laud, P., Pankova, A.: Privacy-preserving frequent itemset mining for sparse
  and dense data. In: Nordic Conference on Secure IT Systems. pp. 139--155.
  Springer (2017)

\bibitem{Liu19}
Liu, C., Duan, H., Zeng, Q., Zhou, M., Lu, F., Cheng, J.: Towards comprehensive
  support for privacy preservation cross-organization business process mining.
  {IEEE} Trans. Services Computing  \textbf{12}(4),  639--653 (2019)

\bibitem{MannhardtKBWM19}
Mannhardt, F., Koschmider, A., Baracaldo, N., Weidlich, M., Michael, J.:
  Privacy-preserving process mining - differential privacy for event logs.
  Business {\&} Information Systems Engineering  \textbf{61}(5),  595--614
  (2019)

\bibitem{mannhardt2018privacy}
Mannhardt, F., Petersen, S.A., Oliveira, M.F.: Privacy challenges for process
  mining in human-centered industrial environments. In: 2018 14th International
  Conference on Intelligent Environments (IE). pp. 64--71. IEEE (2018)

\bibitem{pika2019towards}
Pika, A., Wynn, M.T., Budiono, S., ter Hofstede, A.H., van~der Aalst, W.M.,
  Reijers, H.A.: Towards privacy-preserving process mining in healthcare. In:
  Proc. of the Workshop on Process-Oriented Data Science in Healthcare (PODS4H)
  (2019)

\bibitem{simpda/RafieiWA18}
Rafiei, M., von Waldthausen, L., van~der Aalst, W.M.P.: Ensuring
  confidentiality in process mining. In: Proceedings of the 8th International
  Symposium on Data-driven Process Discovery and Analysis {(SIMPDA} 2018),
  Seville, Spain, December 13-14, 2018. pp. 3--17 (2018)

\bibitem{schulz2004facilitating}
Schulz, K.A., Orlowska, M.E.: Facilitating cross-organisational workflows with
  a workflow view approach. Data \& Knowledge Engineering  \textbf{51}(1),
  109--147 (2004)

\bibitem{shamir1979share}
Shamir, A.: How to share a secret. Communications of the ACM  \textbf{22}(11),
  612--613 (1979)

\bibitem{sweeney2002k}
Sweeney, L.: k-anonymity: A model for protecting privacy. International Journal
  of Uncertainty, Fuzziness and Knowledge-Based Systems  \textbf{10}(05),
  557--570 (2002)

\bibitem{tillem2017mining}
Tillem, G., Erkin, Z., Lagendijk, R.L.: Mining encrypted software logs using
  alpha algorithm. In: SECRYPT. pp. 267--274 (2017)

\bibitem{yao1982protocols}
Yao, A.C.: Protocols for secure computations. In: 23rd annual symposium on
  foundations of computer science (sfcs 1982). pp. 160--164. IEEE (1982)

\bibitem{zeng2013cross}
Zeng, Q., Sun, S.X., Duan, H., Liu, C., Wang, H.: Cross-organizational
  collaborative workflow mining from a multi-source log. Decision support
  systems  \textbf{54}(3),  1280--1301 (2013)

\end{thebibliography}

\end{document}